\documentclass[a4paper,UKenglish,cleveref, autoref, thm-restate]{lipics-v2021}

\sethlcolor{yellow}
\usepackage{xurl}
\usepackage{booktabs}
\usepackage{hhline}
\usepackage{xcolor}
\usepackage{colortbl}
\usepackage{makecell}
\usepackage{textgreek}
\usepackage{siunitx}
\usepackage{stfloats}%
\usepackage{tablefootnote}
\usepackage{bookmark}
\usepackage{etoolbox}
\usepackage{url}
\usepackage{ragged2e}
\usepackage{tikz}
\usepackage{pgfplots}
\pgfplotsset{compat=1.18}
\usepackage{subcaption}
\usepackage{pgf-pie}
\usepackage[most]{tcolorbox}
\newcolumntype{Y}{***{\centering\arraybackslash}X}
\usetikzlibrary{shapes.geometric, arrows.meta, positioning}

\title{Rethinking Automated Program Repair: The Impact of Bug Complexity, Fault Localization, and LLM Cost-efficiency} 

\titlerunning{Rethinking Automated Program Repair} 

\author{Junchi Liu}{Colorado State University, United States}{junchi.liu@colostate.edu}{https://orcid.org/0009-0005-1662-3856}{}

\author{Ali Bigdeli}{Colorado State University, United States}{ali.bigdeli@colostate.edu}{}{}

\author{Roya Daneshi}{Colorado State University, United States}{roya.daneshi@colostate.edu}{}{}

\author{Atu Ambala}{Colorado State University, United States}{atu.ambala@colostate.edu}{}{}

\author{Sudipto Ghosh}{Colorado State University, United States}{Sudipto.Ghosh@colostate.edu}{}{}

\author{Fabio Santos}{Colorado State University, United States}{fabio.deabreusantos@colostate.edu}{https://orcid.org/0000-0001-8069-3158}{}

\authorrunning{J. Liu et al.} 

\Copyright{Junchi Liu, Ali Bigdeli, Roya Daneshi, Atu Ambala, Sudipto Ghosh, and Fabio Santos} 



\category{Technical Track Paper} 

\relatedversion{} 

\supplement{}

\acknowledgements{}

\nolinenumbers 

\EventEditors{Robert Feldt, Maria Paasivaara, Daniel Mendez, Stefan Wagner, and Marvin Mu\~{n}oz Bar\'{o}n}
\EventNoEds{5}
\EventLongTitle{20th International Symposium on Empirical Software Engineering and Measurement (ESEM 2026)}
\EventShortTitle{ESEM 2026}
\EventAcronym{ESEM}
\EventYear{2026}
\EventDate{October 8--9, 2026}
\EventLocation{Munich, Germany}
\EventLogo{}
\SeriesVolume{394}
\ArticleNo{40}

\begin{document}

\maketitle

\begin{abstract}

Background: Software bugs remain a critical challenge in development, necessitating effective Automated Program Repair (APR) techniques. While Large Language Model (LLM)-based APR systems have shown promise, prior studies primarily focus on overall repair effectiveness. The effects of bug complexity, fault localization, reasoning settings, and repair cost-effectiveness remain insufficiently explored.

Aims: This study presents a comprehensive empirical analysis of LLM-based APR, focusing on how repair performance is shaped by bug complexity, fault localization, reasoning settings, and costs.

Method: We construct a curated dataset by collecting algorithmic bugs from AtCoder, a competitive programming platform. We evaluate two APR techniques (ChatRepair and CodeCorrector) using three LLMs (DeepSeek, GPT, and Llama), with multiple model variants and reasoning settings, and examine their performance across diverse levels of bug complexity and localization strategies through a multi-dimensional empirical framework and statistical analysis.

Results: Although structurally complex bugs and imprecise fault localization make repair more challenging, LLM-based APR techniques still achieve competitive repair effectiveness. Imprecise fault localization can substantially enlarge the performance gap between APR techniques. Furthermore, higher-cost LLMs and stronger reasoning settings do not consistently yield better cost-efficiency, revealing a nontrivial trade-off between repair effectiveness and computational cost. We further observe that the impact of reasoning strategies varies considerably across different LLM families, affecting both repair effectiveness and cost-efficiency.

Conclusions: Over 50\% of moderately complex bugs can be repaired by low-cost LLM-based APR techniques. The repair effectiveness gap between APR techniques becomes larger as fault localization becomes less precise. GPT-5 repairs 7 and 39 more complex bugs than DeepSeek-V4-pro and DeepSeek-V3.2, respectively; whereas the total repair cost of DeepSeek-V3.2 across the
non-reasoning and reasoning stages is only approximately 4.6\% and
7.7\% of GPT-5 and DeepSeek-V4-pro, respectively.

\end{abstract}

\section{Introduction}
\label{sec:introduction}

In recent years, the rapid advancement of Generative AI, driven by intense competition among technology companies, has significantly accelerated the development of Large Language Models (LLMs)~\cite{naveed2025comprehensive}. Current LLMs have shown substantial improvements in their ability to understand, reason, and generate programs~\cite{xu2025toward, bahrini2023chatgpt, sengar2025generative}. As a result, they have been increasingly applied to a wide range of software engineering tasks, including code generation, code summarization, and bug detection~\cite{huynh2025large, li2022competition, joel2025survey}.

In the context of Automated Program Repair (APR), LLM-based approaches have demonstrated significant advantages over traditional techniques, achieving strong performance across widely used benchmarks such as Defects4J~\cite{just2014defects4j}, Bugs.jar~\cite{saha2018bugsjar}, and QuixBugs~\cite{lin2017quixbugs}. Notably, many APR techniques achieve near-perfect performance on QuixBugs and IntroClass~\cite{le2015manybugs}, small-scale algorithmic datasets primarily composed of simple, single-hunk bugs, suggesting that these datasets may be insufficiently challenging for evaluating the effectiveness of modern LLM-based APR approaches. Also, as LLMs exhibit strong performance in code understanding and reasoning, recent studies have increasingly explored end-to-end or self-directed APR approaches that reduce reliance on externally provided fault localization (FL) information~\cite{jin2023inferfix, li2025context}. Nevertheless, the impact of the absence of FL on repair effectiveness has not been systematically evaluated in the agent era. To address these unexplored gaps, our study introduces a more challenging algorithmic dataset, featuring bugs with varied complexity levels, and uses the tools of Git~\cite{spinellis2012git}, srcML~\cite{collard2013srcml}, and GumTreeDiff~\cite{falleri2014gumtree} to systematically and automatically categorize bugs based on their complexity, enabling a fine-grained empirical analysis of repair performance across different bug complexity levels as well as under accurate, vague, or no fault localization settings.

Additionally, the reliance on commercial LLM APIs produces non-negligible computational costs~\cite{wang2023cost}. APR tasks often require generating numerous candidate patches~\cite{lin2024one, zhou2024leveraging}, resulting in substantial output token usage, particularly under reasoning-intensive settings~\cite{yue2024large, han2025token}. Consequently, different LLM choices and reasoning settings can lead to significant cost discrepancies. Such costs are often overlooked in prior APR evaluations, despite being critical for practical deployment. In our study, we incorporate cost-efficiency metrics and standardize the unit to \$1 for fair comparison. Specifically, we define Repaired Bugs/\$ and Attempts/\$ as the numbers of successfully repaired bugs and repair attempts divided by the total API cost, respectively, where the cost is calculated from input/output token usage and the corresponding API prices.

We present a comprehensive evaluation of two LLM-based APR techniques across three LLMs, including multiple versions and reasoning settings. We employ multi-dimensional metrics (number of compilable patches, plausible patches, and repaired bugs, $\text{Top}-k$, consistency, and cost-efficiency metrics), and statistical tests to address the following research questions (RQs):

\vspace{1mm}
\noindent \hangindent0.5em \textit{\textbf{RQ1.} How does bug complexity affect the repair performance of LLM-based APR?}

\vspace{1mm}
\noindent \hangindent0.5em\textit{\textbf{RQ2.} To what extent does fault localization influence the effectiveness of LLM-based APR?}

\vspace{1mm}
\noindent \hangindent0.5em\textit{\textbf{RQ3.} How do LLM choices and reasoning settings impact the repair performance–cost trade-off in APR?}
\vspace{1mm}

The results of this study show that:

\begin{itemize}
    \item Repair performance varies substantially across bug complexity, with the simplest complexity bugs being consistently the easiest to fix, as expected, but repair performance remains at a relatively considerable level as complexity increases, suggesting that LLM-based APR approaches maintain a stable reasoning ability even on structurally complex bugs.

    \item FL plays a critical role in improving both repair effectiveness and patch consistency. Although more precise localization generally leads to better repair performance, appropriate APR frameworks can still maintain relatively stable repair effectiveness in settings with vague or missing localization, suggesting potential for practical end-to-end APR scenarios. Also, the impact of FL precision on repair effectiveness decreases as bug complexity increases.

    \item   More costly LLMs and reasoning strategies do not always produce proportional repair improvements; the increased cost should be considered seriously, and reasoning strategies do not consistently yield ideal outcomes, revealing a nontrivial trade-off between performance and efficiency.
\end{itemize}

Our contributions include:
\begin{enumerate}[(i)]
    \item We empirically analyze the impact of bug complexity and fault localization on LLM-based APR performance, and incorporate unit-standardized cost-efficiency metrics to quantify the trade-off between repair effectiveness and computational cost. We also provide an extensible experimentation infrastructure to support future studies, replication efforts, and systematic evaluation of LLM-based APR techniques.

    \item Our study offers novel insights into how repair effectiveness varies with bug complexity and reveals the nontrivial cost implications of different LLM choices and reasoning settings. In particular, by systematically evaluating settings with accurate, vague, and no fault localization, we demonstrate the feasibility and potential of end-to-end LLM-based APR.

    \item Additionally, our results across multiple LLMs and reasoning strategies provide practical guidance for selecting appropriate LLMs under the trade-off between the repair performance and economic cost, highlighting the importance of balancing repair performance and computational cost.
\end{enumerate}

\section{Background and Related Work}\label{sec:relatedwork}

This section provides a relevant review of the literature and highlights the limitations of existing studies.

\subsection{Previous Evaluations of LLM-based APR}

Recent evaluations of APR are primarily based on the Defects4J Benchmark~\cite{just2014defects4j, martinez2017automatic}, which has become the dominant benchmark for APR evaluation. However, heavy reliance on a single benchmark can limit the diversity of bug types and weaken the evaluation of APR approaches' generalizability. Nevertheless, several variants of Defects4J have been proposed, including DEFECTS4J-TRANS~\cite{li2025evaluating}, and MuBench~\cite{ouyang2024benchmarking}. These benchmarks still primarily contain software-engineering–oriented bugs~\cite{renzullo2025automated}. Meanwhile, algorithm-oriented datasets such as QuixBugs~\cite{lin2017quixbugs}, IntroClass, and ManyBugs~\cite{le2015manybugs} have been widely utilized to evaluate the generalizability of APR techniques. However, the difficulty of these datasets is often insufficient to challenge recent LLM-based APR approaches~\cite{xia2024automated}, thereby leading to an overestimation of their effectiveness. In addition, the continued reuse of long-standing benchmark datasets introduces data leakage issues~\cite{renzullo2025automated}, potentially leading to biased evaluation results. Furthermore, Bean et al.~\cite{bean2026measuring} point out that the lack of statistical testing in the majority of benchmarks reduces the reliability of empirical findings.

We collect recent real-world bugs from AtCoder~\cite{atcoder} as an extension of algorithm-oriented datasets. Then, Git, srcML, and GumTreeDiff are utilized to categorize bugs based on complexity. Finally, we conduct an empirical evaluation of LLM-based APR under different bug complexity and FL settings, with a cost-efficiency and statistical analysis.

\subsection{APR Techniques}

APR techniques have advanced significantly with the development of LLMs' code-reasoning and code-generation capabilities~\cite{zubair2025use}. With the success of LLMs in code reasoning and generation~\cite{jiang2026survey}, recent studies have increasingly explored their applications in APR. Xia and Zhang~\cite{xia2024automated} introduced ChatRepair, which iteratively generates and refines patches by leveraging feedback from both failed and successful repair attempts, and has been widely viewed as a representative baseline in subsequent studies. For instance, Yin et al.~\cite{yin2024thinkrepair} proposed ThinkRepair, which uses Chain-of-Thought (CoT) prompting~\cite{wei2022chain} to enable reasoning prior to patch generation. Bouzenia et al.~\cite{bouzenia2025repairagent} introduced RepairAgent, the first autonomous agent-based framework for APR. Li et al.~\cite{li2025context} further proposed CodeCorrector, an end-to-end framework that first analyzes programs to infer repair directions and then generates patches, achieving the best performance on the Defects4J v1.2 dataset. 

In this study, we select ChatRepair as a representative LLM-based APR approach and CodeCorrector as a recent LLM-based method that demonstrates strong performance in prior studies.

\subsection{Evaluation Metrics}

Existing APR evaluation metrics primarily focus on repair effectiveness, including the number of repaired bugs, $\textit{Top}-k$, compilable patches, and plausible patches~\cite{dikici2025advancements}. To further assess patch quality, Dai et al.~\cite{dai2025less} introduced the consistency metric, which measures the extent of changes between patches and buggy programs. However, these metrics fail to capture cost-efficiency, which has become increasingly important in the empirical study of LLM-based APR. Although recent studies have begun to report costs~\cite{xia2024automated, bouzenia2025repairagent, li2025context}, they use inconsistent units, such as cost per bug, cost per successful repair, or cost per attempt, making it difficult to perform fair, direct comparisons across approaches. This lack of a unified cost metric limits the systematic evaluation of performance–cost trade-offs. Thus, we standardize costs to a common monetary unit, enabling a direct trade-off between repair performance and cost.

\section{Research Design}\label{researchdesign}

In this section, we discuss the research methodology.

\subsection{Experimental Infrastructure}
The experimental infrastructure is deployed on computing servers with isolated execution environments to ensure controlled access and protected communication. Shared storage and distributed computing resources are utilized to support large-scale APR experiments across multiple machines, improving the overall experimental efficiency. The experimental infrastructure consists of five components: the dataset, APR techniques, LLMs, additional tools, and Python scripts.  

The dataset is constructed from AtCoder and includes problem descriptions, pairs of incorrect and correct C++ programs, and corresponding test cases (Section~\ref{subsec:dataset}). The tools include the APR tools, along with Git, srcML, and GumTreeDiff. The AtCoder Library (ACL) is included as an external dependency to support compilation. Custom Python scripts are used for data analysis and dataset filtering. All the above components are maintained within our isolated environment.

We utilize multiple LLMs in our study: gpt-5-2025-08-07 (GPT-5)~\cite{openai_gpt5}, gpt-4o-mini-2024-07-18 (GPT-4o-mini)~\cite{openai_gpt4omini}, DeepSeek-V3.2~\cite{deepseek_v32}, DeepSeek-V4-pro~\cite{deepseek_v4}, Llama-3.1-8B~\cite{llama31_8b}, and Llama-3.1-405B~\cite{llama31_405b}. GPT-5 and GPT-4o-mini use the OpenAI API. DeepSeek-V3.2 and DeepSeek-V4-pro use the DeepSeek API. Llama-3.1 is deployed in our experimentation infrastructure.

\input{Tables/research_flow}
\subsection{Experimental Workflow}

Figure~\ref{fig:research_flow} illustrates the overall experimental infrastructure and workflow of our study. We first collect buggy and corrected programs from AtCoder and construct an automated benchmark and evaluation pipeline for LLM-based APR experiments (details are provided in subsection~\ref{subsec:dataset}).

Next, we analyze bug complexity from multiple perspectives. Git-based differencing is utilized to extract hunk-line complexity information, while srcML~\cite{collard2013srcml} and GumTreeDiff~\cite{falleri2014gumtree} are employed to measure structural complexity at the Abstract Syntax Tree (AST) level (details are provided in subsection~\ref{subsec:bug_complexity}).

Based on the constructed benchmark and complexity analysis, we deploy APR frameworks and integrate multiple LLM APIs with different reasoning settings to conduct automated repair experiments under FL settings (details are provided in subsection~\ref{subsec:bug_localization_and_apr_settings}).

Finally, we perform a comprehensive empirical evaluation to investigate the impact of bug complexity, fault localization, and cost-efficiency trade-offs on the performance of LLM-based APR techniques. The corresponding findings and experimental analyses are presented in section~\ref{results}.

\subsection{Dataset}\label{subsec:dataset}

Existing APR benchmarks have been widely used, raising concerns about data contamination in the evaluation of LLM-based APR. To mitigate this risk, we construct our dataset from the recent AtCoder submissions rather than relying on existing benchmarks. We focus on C++ programs because comparing FL settings and bug complexity across programming languages with different characteristics could introduce additional bias. This design enables a more controlled evaluation of LLM-based APR.

In this study, we use AtCoder, a Japanese competitive programming platform, and collect $1{,}914$ C++ source code submissions in October, 2025. Our data comes from regular contests, which typically include algorithmically challenging problems designed to assess intermediate-to-advanced programming skills. We pair every incorrect program and correct program from the same participant as a pair of a bug and a correct patch (bug-patch pair). For each bug-patch pair, we use the \texttt{git diff} command to generate a diff file that captures the corresponding code changes.

Before generating diff files using Git, we remove all comments and redundant blank lines. The diff files reflect line-level code changes, where `-` denotes deleted (buggy) lines $N_{-}$, and `+` denotes added (patch) lines $N_{+}$. For each hunk, the number of fixed lines is defined as $N_{\text{fixed}}=\max(N_{+},N_{-})$.

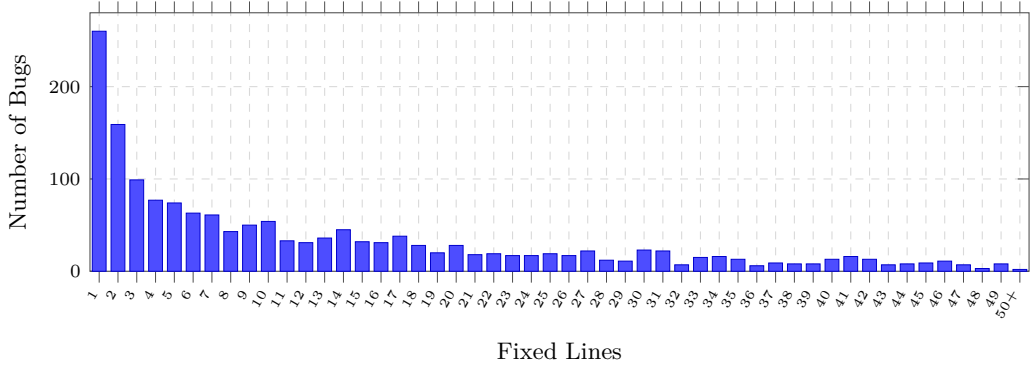
\begin{figure*}[t]
    \centering
    \begin{tikzpicture}
    \begin{axis}[
        width=1\textwidth,
        height=5.0cm,
        ybar,
        bar width=5.2pt,
        ymin=0,
        ymax=280,
        xlabel={Fixed Lines},
        ylabel={Number of Bugs},
        xtick={1,2,...,50},
        xticklabels={
            1,2,3,4,5,6,7,8,9,10,
            11,12,13,14,15,16,17,18,19,20,
            21,22,23,24,25,26,27,28,29,30,
            31,32,33,34,35,36,37,38,39,40,
            41,42,43,44,45,46,47,48,49,50+
        },
        x tick label style={font=\tiny, rotate=60, anchor=east},
        y tick label style={font=\scriptsize},
        label style={font=\small},
        grid=major,
        grid style={dashed, gray!30},
        axis line style={black!70},
        tick style={black!70},
        enlarge x limits=0.01,
    ]

    \addplot[
        fill=blue!70,
        draw=blue!80!black,
        line width=0.25pt
    ] coordinates {
        (1,260)
        (2,159)
        (3,99)
        (4,77)
        (5,74)
        (6,63)
        (7,61)
        (8,43)
        (9,50)
        (10,54)
        (11,33)
        (12,31)
        (13,36)
        (14,45)
        (15,32)
        (16,31)
        (17,38)
        (18,28)
        (19,20)
        (20,28)
        (21,18)
        (22,19)
        (23,17)
        (24,17)
        (25,19)
        (26,17)
        (27,22)
        (28,12)
        (29,11)
        (30,23)
        (31,22)
        (32,7)
        (33,15)
        (34,16)
        (35,13)
        (36,6)
        (37,9)
        (38,8)
        (39,8)
        (40,13)
        (41,16)
        (42,13)
        (43,7)
        (44,8)
        (45,9)
        (46,11)
        (47,7)
        (48,3)
        (49,8)
        (50,2)
};

    \end{axis}
    \end{tikzpicture}
    \vspace{-1.6mm}
    \caption{Distribution of Fixed Size Across Raw Data}
    \vspace{-1.6mm}
    \label{fig:distribution_of_bug_fix_sizes}
\end{figure*}

The distribution of bug-fix sizes across the original 1,914 C++ files is shown in Figure~\ref{fig:distribution_of_bug_fix_sizes}. Based on preliminary experiments comparing different fixed line thresholds on samples drawn from the full dataset, we observe that APR becomes considerably more difficult and computationally expensive as the number of fixed lines increases. Considering both repair feasibility and computational cost, we set a threshold of five fixed lines to focus on small-scope program repair. Finally, 640 bugs remain for our empirical study.

\subsection{Bug Complexity}\label{subsec:bug_complexity}

\textbf{Buggy hunk-line-level complexity}: The diff files generated by Git not only identify buggy and patched lines but also organize code changes into hunks. We use Git to generate diff files for every bug-patch pair. Based on the number of hunks and the number of fixed lines within each hunk, we categorize bugs into four categories: Single-Hunk-Single-Line (SHSL), Single-Hunk-Multi-Line (SHML), Multi-Hunk-Single-Line (MHSL), and Multi-Hunk-Multi-Line (MHML). The categorization rule is defined in ~\autoref{eq:buggy-hunk-and-line-level-complexity}. The number of hunks can be obtained directly from Git diff files. If any hunk contains more than one fixed line, the bug is classified as a multi-line bug.

\vspace{-3mm}
\begin{equation}
\begin{cases}
\text{SHSL}, & H = 1 \land \forall h_i \in H,\; L(h_i)=1 \\[2pt]

\text{SHML}, & H = 1 \land \exists h_i \in H,\; L(h_i)>1 \\[2pt]

\text{MHSL}, & H > 1 \land \forall h_i \in H,\; L(h_i)=1 \\[2pt]

\text{MHML}, & H > 1 \land \exists h_i \in H,\; L(h_i)>1
\end{cases}
\label{eq:buggy-hunk-and-line-level-complexity}
\end{equation}
\vspace{-3mm}

where $H$ denotes the total number of hunks in a bug, $h_i$ represents the $i$-th hunk ($1 \leq i \leq H$), and $L(h_i)$ denotes the number of fixed lines in hunk $h_i$.

\textbf{AST operation-type complexity}: We employ srcML to construct XML-based structural representations of programs~\cite{collard2013srcml} and further utilize GumTreeDiff to extract AST-level edit operations between buggy and corrected program pairs~\cite{falleri2014gumtree}. Unlike hunk-line-level complexity, this metric measures structural differences at the AST node level, providing a finer-grained view of bug complexity.

\begin{figure}[t]
    \centering
    \includegraphics[width=0.95\columnwidth]{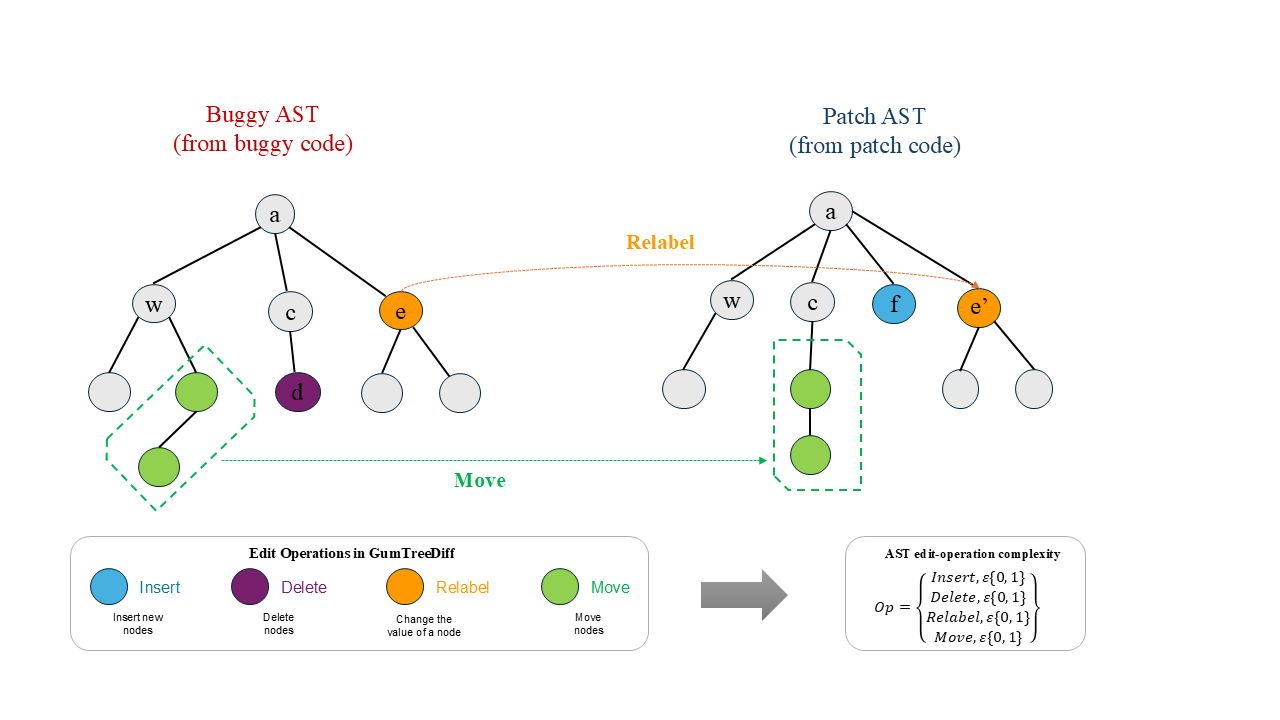}
    \vspace{-5mm}
    \caption{AST operation-type complexity.}
    \vspace{-5mm}
    \label{fig:ASToperation}
\end{figure}

The AST edit distance~\cite{song2024revisiting} between a buggy and a correct program is defined by four edit operations: insert, delete, relabel, and move. In our study, bugs are further categorized based on the diversity of edit operations required for repair. \autoref{fig:ASToperation} illustrates how the AST edit-operation-type complexity is categorized.

\autoref{tab:hunk_line_bug_count} and \autoref{tab:ast_operation_bug_count} present the number of bugs in each type under the hunk-line-level and AST operation-type complexity, respectively. We conduct empirical evaluations to analyze how bug complexity affects LLM-based APR performance.

\begin{table*}[t]
\centering
\footnotesize
\setlength{\tabcolsep}{5pt}
\renewcommand{\arraystretch}{1.2}

\begin{minipage}[t]{0.47\textwidth}
\centering
\caption{Hunk-line complexity.}
\label{tab:hunk_line_bug_count}

\begin{tabularx}{\textwidth}{>{\raggedright\arraybackslash}X c}
\toprule
\textbf{Bug Type} & \textbf{\# Bugs} \\
\midrule
Single Hunk Single Line & 251 \\
Single Hunk Multi Line  & 226 \\
Multi Hunk Single Line  & 67 \\
Multi Hunk Multi Line   & 96 \\
\midrule
Total   & 640 \\
\bottomrule
\end{tabularx}
\end{minipage}
\hfill
\begin{minipage}[t]{0.47\textwidth}
\centering
\caption{AST operation-type complexity}
\label{tab:ast_operation_bug_count}

\begin{tabularx}{\textwidth}{>{\raggedright\arraybackslash}X c}
\toprule
\textbf{Bug Type} & \textbf{\# Bugs} \\
\midrule
Single Operation & 278 \\
Two Operations   & 158 \\
Three Operations & 121 \\
Four Operations  & 83 \\
\midrule
Total   & 640 \\
\bottomrule
\end{tabularx}
\end{minipage}

\end{table*}

\subsection{Fault Localization And APR Settings}\label{subsec:bug_localization_and_apr_settings}

\begin{figure}[t]
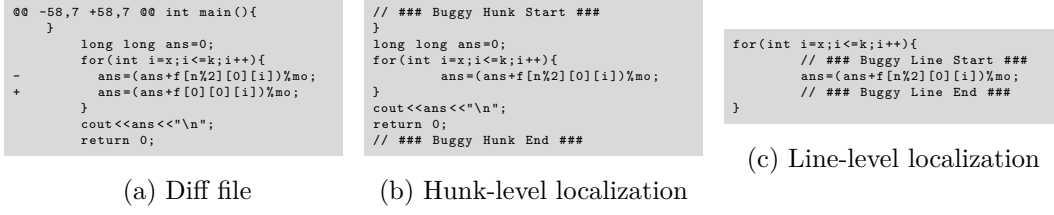

\centering

\begin{minipage}{0.32\columnwidth}
\begin{lstlisting}[basicstyle=\ttfamily\tiny]
@@ -58,7 +58,7 @@ int main(){
    }
 	long long ans=0;
 	for(int i=x;i<=k;i++){
-	  ans=(ans+f[n%2][0][i])%mo;
+	  ans=(ans+f[0][0][i])%mo;
 	}
 	cout<<ans<<"\n";
 	return 0;
\end{lstlisting}
\hspace{1mm}
\centerline{(a) Diff file}
\end{minipage}
\hfill
\begin{minipage}{0.32\columnwidth}
\begin{lstlisting}[basicstyle=\ttfamily\tiny]
// ### Buggy Hunk Start ###
}
long long ans=0;
for(int i=x;i<=k;i++){
	ans=(ans+f[n%2][0][i])%mo;
}
cout<<ans<<"\n";
return 0;
// ### Buggy Hunk End ###
\end{lstlisting}
\centerline{(b) Hunk-level localization}
\end{minipage}
\hfill
\begin{minipage}{0.32\columnwidth}
\begin{lstlisting}[basicstyle=\ttfamily\tiny]
for(int i=x;i<=k;i++){
	// ### Buggy Line Start ###
	ans=(ans+f[n%2][0][i])%mo;
	// ### Buggy Line End ###
}
\end{lstlisting}
\centerline{(c) Line-level localization}
\end{minipage}

\caption{Fault Localization Strategy Examples}
\label{fig:buglocalization}
\end{figure}

We use three FL strategies to explore their impact on LLM-based APR: buggy line-level FL, buggy hunk-level FL, and no FL. \autoref{fig:buglocalization} presents the examples of line-level FL and hunk-level FL according to the diff file.

Different LLMs provide different mechanisms for controlling reasoning behavior. (a) DeepSeek-V3.2 provides two model variants: \textit{DeepSeek-V3.2-chat} and \textit{DeepSeek-V3.2-reasoner}. For DeepSeek-V4-pro, the reasoning mode can be enabled or disabled through the \texttt{thinking\_type} parameter in the DeepSeek API. Additionally, the \texttt{reasoning\_effort} parameter can be configured as \texttt{high} or \texttt{max}. (b) GPT models do not provide separate chat or reasoner variants; instead, reasoning intensity can only be controlled through the \texttt{reasoning.effort} parameter, with settings such as \texttt{medium}, \texttt{high}, and \texttt{xhigh}. For Llama models, there is currently no explicit mechanism to directly control reasoning settings.

Furthermore, the temperature for all LLMs is set to 1.0~\cite{renze2024effect} for all experiments. For ChatRepair, we set the maximum number of conversation rounds to 3~\cite{xia2024automated}. For both ChatRepair and CodeCorrector, we provide the problem background, the buggy code, the error message, and at most 3 failed test cases. We retain the original prompts reported in the corresponding studies~\cite{xia2024automated, li2025context}, and we do not perform model-specific parameter tuning, allowing the comparison to focus on the influence of the APR frameworks, FL strategies, and LLM effectiveness.

\subsection{Experimental Design}\label{subsec:experimental_design}

To balance experimental comprehensiveness and computational cost, we adopt a three-stage experimental design.

First, we evaluate two APR tools with three different low-cost LLMs on the complete set of $640$ bugs under line-level FL: \textit{DeepSeek-V3.2-chat}, \textit{GPT-4o-mini}, and \textit{Llama-3.1-8B}. The maximum number of repair rounds is set to 10 as in~\cite{palavalli2024using}. We conduct a comprehensive empirical analysis of various APR techniques and LLM combinations with respect to repair effectiveness and cost-efficiency.

Next, we select the best-performing LLM from the first stage and further evaluate it on all $640$ bugs at line-level, hunk-level, and no-FL. The results are then compared across three FL strategies to analyze the impact of different FL precision on two LLM-based APR techniques. For the first two stages, reasoning mode is disabled across all LLMs, allowing us to identify bugs that can already be repaired under non-reasoning configurations before evaluating more computationally expensive reasoning-enabled models.

To further investigate the repair effectiveness and cost-efficiency of advanced LLMs and reasoning settings under more challenging repair scenarios, we adopt a filtering strategy to focus the evaluation on bugs with moderate repair difficulty while controlling the computational cost of evaluating advanced LLMs. Specifically, we exclude SHSL bugs, single-operation-type bugs, and four-operation-type bugs. In addition, bugs that can already be successfully repaired by at least five APR techniques--LLM combinations in the first-stage experiments are removed. Consequently, the remaining benchmark focuses on bugs with moderate structural complexity and lower repairability under low-cost APR configurations. After filtering, $211$ bugs remain for empirical analysis. We further select the better-performing APR technique from the previous stages as the representative APR method for subsequent experiments.

In this stage, we evaluate four advanced LLMs: \textit{DeepSeek-V4-pro}, \textit{DeepSeek-V3.2}, \textit{GPT-5}, and \textit{Llama-3.1-405B}. The maximum number of repair rounds is increased to $13$. During the first $10$ rounds, reasoning mode is disabled. If a bug remains unfixed, reasoning mode is enabled for the final three rounds. For \textit{GPT-5}, \textit{DeepSeek-V3.2}, and \textit{DeepSeek-V4-pro}, the reasoning effort is configured as \texttt{high}.

\textbf{Experimental Token Consumption.}
To facilitate the reproduction of our experiments, we report the total input and output tokens consumed by each evaluated LLM that incurred API costs. DeepSeek-V3.2 consumed 116.5M input tokens and 42.7M output tokens. GPT-4o-mini used 49.5M input tokens and 14.4M output tokens. DeepSeek-V4-pro consumed 6.3M input tokens and 9.9M output tokens, GPT-5 consumed 3.4M input tokens and 7.7M output tokens, and Llama-3.1-405B consumed 6.3M input tokens and 1.7M output tokens.

\textbf{API Response Time.}
We also report the approximate response time per repair request. Without reasoning mode, all evaluated LLMs require approximately 1--3 minutes per request. With reasoning enabled, DeepSeek-V3.2 and DeepSeek-V4-pro required approximately 40--60 minutes per request, while GPT-5 required about 30 minutes.

\section{Results and Empirical Analysis}
\label{results}

In this section, we show the results of our empirical experiment.

\subsection{APR Performance With Different LLMs}
\label{aprperformancewithdifferentllms}

\begin{table*}[t]
\caption{Comparison of repair effectiveness between CodeCorrector and ChatRepair frameworks using three base LLMs on the full benchmark.}
\label{tab:step1_result1}
\centering
\footnotesize
\setlength{\tabcolsep}{2pt}
\renewcommand{\arraystretch}{1.2}

\begin{tabularx}{\textwidth}{l *{6}{>{\centering\arraybackslash}X}} 
\toprule
\multirow{2}{*}{} & \multicolumn{3}{c}{\textbf{CodeCorrector}} & \multicolumn{3}{c}{\textbf{ChatRepair}} \\ 
\cmidrule(lr){2-4} \cmidrule(lr){5-7}
 & \textbf{DeepSeek-V3.2-chat} & \textbf{GPT-4o-mini} & \textbf{Llama-3.1-8B} & \textbf{DeepSeek-V3.2-chat} & \textbf{GPT-4o-mini} & \textbf{Llama-3.1-8B} \\ \midrule
Repaired Bugs  & 295/640 & 126/640  & 11/640  & 292/640 & 140/640  & 50/640  \\
Attempts & 4003 & 5373  & 6305  & 6400 & 6400  & 6400  \\
Unique Patches & 3117(77.9\%) & 4425(82.4\%)  & 5833(92.5\%)  & 2623(41.0\%) & 3774(59.0\%)  & 5111(80.0\%)  \\
Compilable Patches   & 2830(70.7\%) & 3899(72.6\%) & 1466(23.3\%) & 2523(39.4\%) & 3511(54.9\%)  & 3122(48.8\%) \\
Plausible Patches   & 295 & 126 & 11 & 421 & 247  & 76 \\
Repaired Bugs / \$  & 47.62 & 17.86 & - & 71.43 & 20.41  & -  \\
Attempts / \$ & 646.17 & 761.48 & - & 1565.56 & 932.94 & - \\ \bottomrule
\end{tabularx}
\end{table*}
\autoref{tab:step1_result1} presents the repair performance of CodeCorrector and ChatRepair with three base LLMs under the non-reasoning setting. In this experiment, line-level FL is provided within the incorrect programs, and the maximum repair round is set to 10. CodeCorrector stops after generating the first plausible patch, whereas ChatRepair continues generating candidate patches.

Overall, DeepSeek-V3.2-chat achieves substantially stronger repair effectiveness than GPT-4o-mini across both APR frameworks. When integrated with DeepSeek-V3.2-chat, both CodeCorrector and ChatRepair repair more than twice as many bugs as they do with GPT-4o-mini (295, 292 vs. 126, 140), whereas Llama-3.1-8B demonstrates very limited repair behavior: only 11 bugs were corrected with CodeCorrector and 50 with ChatRepair.

In addition, DeepSeek-V3.2-chat exhibits noticeably better cost-efficiency. DeepSeek-V3.2-chat repairs $47{.}62$ bugs per dollar under CodeCorrector, compared with only 17.86 bugs per dollar for GPT-4o-mini, representing approximately $2.7 \times$ higher cost-efficiency. Similarly, under ChatRepair, DeepSeek achieves $71{.}43$ repaired bugs per dollar, whereas GPT-4o-mini repairs only $20{.}41$ bugs per dollar, corresponding to approximately $3{.}5\times$ higher cost-efficiency. 

\textbf{\textit{Finding 1:}} \textit{Among relatively low-cost LLMs, DeepSeek-V3.2-chat demonstrates the strongest overall APR effectiveness, achieving significantly better repair results and cost-efficiency than GPT-4o-mini, whereas Llama-3.1-8B exhibits limited repair effectiveness despite its negligible inference cost.}

\begin{table*}[t]
\centering
\caption{Pairwise 2$\times$2 contingency tables for chi-square tests.}
\label{tab:pairwise_chisquare}

\footnotesize
\setlength{\tabcolsep}{3.8pt}
\renewcommand{\arraystretch}{1.1}


\textbf{(a) Hunk-line Complexity Comparisons}

\vspace{1mm}

\begin{tabular}{c@{\hspace{0.25cm}}c@{\hspace{0.25cm}}c}

{\scriptsize \textbf{SHSL vs SHML}} &
{\scriptsize \textbf{SHSL vs MHSL}} &
{\scriptsize \textbf{SHSL vs MHML}} \\[-0.5mm]

\begin{tabular}{lcc}
\toprule
Type & Fixed & Unfixed \\
\midrule
SHSL & 197(78.5\%) & 54 \\
SHML & 113(50.0\%) & 113 \\
\bottomrule
\end{tabular}
&
\begin{tabular}{lcc}
\toprule
Type & Fixed & Unfixed \\
\midrule
SHSL & 197(78.5\%) & 54 \\
MHSL & 39(58.2\%) & 28 \\
\bottomrule
\end{tabular}
&
\begin{tabular}{lcc}
\toprule
Type & Fixed & Unfixed \\
\midrule
SHSL & 197(78.5\%) & 54 \\
MHML & 48(50.0\%) & 48 \\
\bottomrule
\end{tabular}

\\[2ex]

{\scriptsize \textbf{SHML vs MHSL}} &
{\scriptsize \textbf{SHML vs MHML}} &
{\scriptsize \textbf{MHSL vs MHML}} \\

\begin{tabular}{lcc}
\toprule
Type & Fixed & Unfixed \\
\midrule
SHML & 113(50.0\%) & 113 \\
MHSL & 39(58.2\%) & 28 \\
\bottomrule
\end{tabular}
&
\begin{tabular}{lcc}
\toprule
Type & Fixed & Unfixed \\
\midrule
SHML & 113(50.0\%) & 113 \\
MHML & 48(50.0\%) & 48 \\
\bottomrule
\end{tabular}
&
\begin{tabular}{lcc}
\toprule
Type & Fixed & Unfixed \\
\midrule
MHSL & 39(58.2\%) & 28 \\
MHML & 48(50.0\%) & 48 \\
\bottomrule
\end{tabular}

\end{tabular}

\vspace{3mm}


\textbf{(b) AST Operation Complexity Comparisons}

\vspace{1mm}

\begin{tabular}{c@{\hspace{0.25cm}}c@{\hspace{0.25cm}}c}

{\scriptsize \textbf{Single vs Two}} &
{\scriptsize \textbf{Single vs Three}} &
{\scriptsize \textbf{Single vs Four}} \\[-0.5mm]

\begin{tabular}{lcc}
\toprule
Type & Fixed & Unfixed \\
\midrule
Single & 209(75.2\%) & 69 \\
Two & 90(57.0\%) & 68 \\
\bottomrule
\end{tabular}
&
\begin{tabular}{lcc}
\toprule
Type & Fixed & Unfixed \\
\midrule
Single & 209(75.2\%) & 69 \\
Three & 66(54.5\%) & 55 \\
\bottomrule
\end{tabular}
&
\begin{tabular}{lcc}
\toprule
Type & Fixed & Unfixed \\
\midrule
Single & 209(75.2\%) & 69 \\
Four & 32(38.6\%) & 51 \\
\bottomrule
\end{tabular}

\\[2ex]

{\scriptsize \textbf{Two vs Three}} &
{\scriptsize \textbf{Two vs Four}} &
{\scriptsize \textbf{Three vs Four}} \\

\begin{tabular}{lcc}
\toprule
Type & Fixed & Unfixed \\
\midrule
Two & 90(57.0\%) & 68 \\
Three & 66(54.5\%) & 55 \\
\bottomrule
\end{tabular}
&
\begin{tabular}{lcc}
\toprule
Type & Fixed & Unfixed \\
\midrule
Two & 90(57.0\%) & 68 \\
Four & 32(38.6\%) & 51 \\
\bottomrule
\end{tabular}
&
\begin{tabular}{lcc}
\toprule
Type & Fixed & Unfixed \\
\midrule
Three & 66(54.5\%) & 55 \\
Four & 32(38.6\%) & 51 \\
\bottomrule
\end{tabular}

\end{tabular}

\end{table*}
\begin{table*}[t]
\centering
\caption{Pairwise Pearson Chi-square test results across different bug complexity categories.}
\label{tab:pairwise_chi_square}

\scriptsize
\setlength{\tabcolsep}{2.5pt}
\renewcommand{\arraystretch}{1.0}

\begin{minipage}{0.47\textwidth}
\centering

\begin{tabular}{lcccc}
\toprule
 & SHSL & SHML & MHSL & MHML \\
\midrule
SHSL & -- & \cellcolor{red!40}$p<0.0001$ & \cellcolor{red!30}$p=0.0013$ & \cellcolor{red!40}$p<0.0001$ \\
SHML & -- & -- & \cellcolor{red!7}$p=0.30$ & $p=1.000$ \\
MHSL & -- & -- & -- & \cellcolor{red!3}$p=0.38$ \\
MHML & -- & -- & -- & -- \\
\bottomrule
\end{tabular}

\vspace{1mm}
(a) Hunk-line complexity
\end{minipage}
\hfill
\begin{minipage}{0.47\textwidth}
\centering

\begin{tabular}{lcccc}
\toprule
 & 1-op & 2-op & 3-op & 4-op \\
\midrule
1-op & -- & \cellcolor{red!40}$p=0.0001$ & \cellcolor{red!40}$p<0.0001$ & \cellcolor{red!40}$p<0.0001$ \\
2-op & -- & -- & \cellcolor{red!1}$p=0.78$ & \cellcolor{red!20}$p=0.010$ \\
3-op & -- & -- & -- & \cellcolor{red!10}$p=0.035$ \\
4-op & -- & -- & -- & -- \\
\bottomrule
\end{tabular}

(b) AST operation type complexity
\end{minipage}

\vspace{1mm}
\scriptsize
Darker cells indicate smaller p-values.
\end{table*}

\textbf{Bug Complexity:} We then analyze how bug complexity affects overall repairability, where a bug is considered repaired if any evaluated APR technique and LLM configuration can fix it. \autoref{tab:pairwise_chisquare} and \autoref{tab:pairwise_chi_square} present the pairwise $2\times2$ contingency tables and Pearson chi-square test results for the numbers of fixed and unfixed bugs across different bug complexity categories. Pairwise chi-square tests are conducted to assess whether differences in repair rates across complexity categories are statistically significant. Following~\cite{mchugh2013chi}, we consider differences statistically significant when $p < 0.05$, while $p < 0.005$ indicates sufficient evidence against the null hypothesis.

SHSL and single-operation-type bugs exhibit the lowest structural complexity and consequently achieve the highest repair rates among all categories, reaching 78.5\% and 75.2\%, respectively. Furthermore, all pairwise comparisons between these two categories and the remaining complexity categories yield $p < 0.005$, providing strong statistical evidence that SHSL and single-operation-type bugs are substantially easier for APR techniques to repair. However, the SHML, MHSL, MHML, two-operation, and three-operation bugs also achieve repair rates above 50\%.

For hunk-line-level complexity, no statistically significant difference in repair is observed among multi-line bug categories ($p > 0.05$). In contrast, although no statistically significant difference is observed between two-operation-type and three-operation-type bugs ($p = 0.78$), four-operation-type bugs exhibit significantly lower repair effectiveness than other operation-type categories ($p < 0.05$). Thus, AST operation complexity further distinguishes highly complex bugs, with statistical analysis showing that four-operation bugs exhibit the most complex repair structure.

\textbf{\textit{Finding 2:}} \textit{SHSL and single-operation-type bugs are the easiest to repair, whereas four-operation-type bugs remain the most challenging. The repair rates of moderate-complexity categories exceed 50\%, indicating that LLM-based APR techniques maintain strong repair effectiveness. Although four-operation-type bugs exhibit lower repair performance, their repair rates remain considerable. These results suggest that low-cost LLM-based APR techniques retain substantial repair effectiveness even for structurally complex bugs.}

\begin{tcolorbox}[
    colback=gray!15,
    colframe=gray!15,
    boxrule=0pt,
    arc=2mm,
    left=2mm,
    right=2mm,
    top=1mm,
    bottom=1mm
]
\textbf{RQ1 Summary.}
 Bug complexity significantly affects repair effectiveness. Chi-square test results suggest three approximate levels of repair difficulty across different bug complexity categories. Overall, repair effectiveness decreases with increasing bug complexity, while the simplest bugs remain substantially easier to repair. Although four-operation bugs are the hardest to repair, APR techniques still maintain considerable repair capability for structurally complex bugs. Over 50\% of moderate-complexity bugs can still be repaired successfully. These results demonstrate the potential of low-cost LLM-based APR techniques for handling challenging real-world repair scenarios.
\end{tcolorbox}

\subsection{Fault Localization Evaluation}
\label{buglocalizationeva}
Given DeepSeek-V3.2-chat's best performance in the previous experiment, we further evaluate the impact of different FL strategies on APR performance using this LLM with both APR techniques. In this experiment, the maximum repair round is still set to 10.

\autoref{tab:step2_result1} shows that CodeCorrector and ChatRepair achieve nearly identical repair effectiveness under line-level FL, with CodeCorrector even repairing three more bugs. However, when only hunk-level FL or no FL is provided, ChatRepair demonstrates consistently greater stability and generalizability. Under hunk-level FL, ChatRepair repairs $58$ more bugs than CodeCorrector, and this gap further increases to $93$ bugs when FL is unavailable. 

\textbf{\textit{Finding 3:}} \textit{The performance gap between APR techniques increases markedly as FL becomes less precise, indicating that repair stability and generalizability are critical for real-world APR scenarios. Different APR techniques may achieve similar repair rates under precise line-level FL, but their repair rates diverge substantially under imprecise FL settings.}

\begin{table*}[t]
\caption{Comparison of repair effectiveness between CodeCorrector and ChatRepair frameworks using three Fault Localization strategies on the full benchmark.}
\label{tab:step2_result1}
\centering
\footnotesize
\setlength{\tabcolsep}{2pt}
\renewcommand{\arraystretch}{1.2}

\begin{tabularx}{\textwidth}{l *{6}{>{\centering\arraybackslash}X}} 
\toprule
\multirow{2}{*}{} & \multicolumn{3}{c}{\textbf{CodeCorrector}} & \multicolumn{3}{c}{\textbf{ChatRepair}} \\ 
\cmidrule(lr){2-4} \cmidrule(lr){5-7}
 & \textbf{line-level} & \textbf{hunk-level} & \textbf{w/o FL} & \textbf{line-level} & \textbf{hunk-level} & \textbf{w/o FL} \\ \midrule
Repaired Bugs  & 295/640 & 179/640  & 112/640  & 292/640 & 237/640  & 205/640  \\
Attempts & 4003 & 4986  & 5480  & 6400 & 6400  & 6400  \\
Unique Patches & 3117(77.9\%) & 4346(87.2\%)  & 4012(73.2\%)  & 2623(41.0\%) & 4741(74.1\%)  & 3261(51.0\%)  \\
Compilable Patches   & 2830(70.7\%) & 3503(70.3\%) & 3756(68.5\%) & 2523(39.4\%) & 4073(63.6\%)  & 3174(49.6\%) \\
Plausible Patches   & 295 & 179 & 112 & 421 & 542  & 299 \\
Repaired Bugs / \$  & 47.62 & 16.60 & 18.87 & 71.43 & 24.41  & 47.62  \\
Attempts / \$ & 646.17 & 462.52 & 923.18 & 1565.56 & 659.13 & 1486.64 \\ \bottomrule
\end{tabularx}
\end{table*}

We further introduce the \textit{code consistency rate} (CCR) to measure how well repaired programs preserve the original code's structure. $\boldsymbol{\mathrm{CCR}=r/k}$ defines the consistency metric, where $k$ denotes the total number of code lines in the repaired program, and $r$ denotes the number of preserved code lines after repair.

\begin{table}[t]
\caption{Comparison of repair consistency under different FL strategies.}
\label{consistency}
\centering
\footnotesize
\setlength{\tabcolsep}{3pt}
\renewcommand{\arraystretch}{1.2}

\begin{tabularx}{0.6\textwidth}{
l
>{\centering\arraybackslash}X
>{\centering\arraybackslash}X
}
\toprule

FL Strategy & \textbf{CodeCorrector} & \textbf{ChatRepair} \\

\midrule

Line-level FL   & 0.85 & 0.92 \\
Hunk-level FL   & 0.75 & 0.83 \\
w/o FL          & 0.54 & 0.85 \\

\bottomrule
\end{tabularx}

\end{table}

According to \autoref{consistency}, line-level FL achieves the highest patch consistency across all FL settings for both APR techniques. ChatRepair generally generates patches with higher structural consistency than CodeCorrector, particularly without FL (0.85 vs. 0.54).

\textbf{\textit{Finding 4:}} \textit{FL not only improves repair effectiveness but also helps generate patches that remain more structurally consistent with the original programs. The design of APR techniques also significantly impacts the structural consistency of generated patches.}

\begin{figure*}[t]
\centering

\begin{minipage}{0.49\textwidth}
\centering
\begin{tikzpicture}
\begin{axis}[
    legend style={
        at={(100,0.97)},
        anchor=north west,
        font=\tiny,
        draw=none,
        fill=white,
        fill opacity=0.7,
        text opacity=1,
        row sep=-1pt
    },
    legend cell align=left,
    width=\textwidth,
    height=5.5cm,
    ymin=0, ymax=90,
    ylabel={Repair Rate (\%)},
    xlabel={Hunk-line Complexity},
    symbolic x coords={SHSL,SHML,MHSL,MHML},
    xtick=data,
    xlabel style={font=\small},
    ylabel style={font=\small},
    xticklabel style={font=\small},
    yticklabel style={font=\small},
    ytick={0,20,40,60,80},
    yticklabels={0\%,20\%,40\%,60\%,80\%},
    grid=both,
    major grid style={gray!20},
    legend style={
    at={(0.97,0.97)},
    anchor=north east,
    font=\tiny,
    draw=none,
    fill=white,
    fill opacity=0.7,
    text opacity=1,
    row sep=-1pt
},
legend cell align=right,
    nodes near coords,
    every node near coord/.append style={font=\small, yshift=5pt}
]

\addplot[
    color=blue!70,
    line width=1pt,
    mark=*,
    mark size=1pt,
    nodes near coords,
    every node near coord/.append style={
        font=\fontsize{6.5}{7.5}\selectfont,
        yshift=-2pt,
        xshift=1pt
    }
]
coordinates {(SHSL,74.1)(SHML,48.2)(MHSL,52.2)(MHML,49.0)};
\addlegendentry{Line-Level}
\addplot[
    color=orange!70,
    line width=1pt,
    mark=*,
    mark size=1pt,
    nodes near coords,
    every node near coord/.append style={
        font=\fontsize{6.5}{7.5}\selectfont,
        yshift=-3pt
    }
]
coordinates {(SHSL,59.0)(SHML,38.5)(MHSL,37.3)(MHML,32.3)};
\addlegendentry{Hunk-Level}
\addplot[
    color=red!70,
    line width=1pt,
    mark=*,
    mark size=1pt,
    nodes near coords,
    every node near coord/.append style={
        font=\fontsize{6.5}{7.5}\selectfont,
        yshift=-20pt
    }
]
coordinates {(SHSL,47.4)(SHML,30.1)(MHSL,25.4)(MHML,30.2)};
\addlegendentry{w/o FL}
\end{axis}
\end{tikzpicture}
\end{minipage}
\hfill
\begin{minipage}{0.49\textwidth}
\centering
\begin{tikzpicture}
\begin{axis}[
    legend style={
        at={(100,0.97)},
        anchor=north west,
        font=\tiny,
        draw=none,
        fill=white,
        fill opacity=0.7,
        text opacity=1,
        row sep=-1pt
    },
    legend cell align=left,
    width=\textwidth,
    height=5.5cm,
    ymin=0, ymax=90,
    ylabel={Repair Rate (\%)},
    xlabel={AST-operation Type Complexity},
    symbolic x coords={Single-o,Two-o,Three-o,Four-o},
    xtick=data,
    xlabel style={font=\small},
    ylabel style={font=\small},
    xticklabel style={font=\small},
    yticklabel style={font=\small},
    ytick={0,20,40,60,80},
    yticklabels={0\%,20\%,40\%,60\%,80\%},
    grid=both,
    major grid style={gray!20},
    legend style={
    at={(0.97,0.97)},
    anchor=north east,
    font=\tiny,
    draw=none,
    fill=white,
    fill opacity=0.7,
    text opacity=1,
    row sep=-1pt
},
legend cell align=right,
    nodes near coords,
    every node near coord/.append style={font=\small, yshift=5pt}
]
\addplot[
    color=blue!70,
    line width=1pt,
    mark=*,
    mark size=1pt,
    nodes near coords,
    every node near coord/.append style={
        font=\fontsize{6.5}{7.5}\selectfont,
        yshift=-3pt
    }
]
coordinates {(Single-o,70.9)(Two-o,53.8)(Three-o,53.7)(Four-o, 36.1)};
\addlegendentry{Line-Level}

\addplot[
    color=orange!70,
    line width=1pt,
    mark=*,
    mark size=1pt,
    nodes near coords,
    every node near coord/.append style={
        font=\fontsize{6.5}{7.5}\selectfont,
        yshift=-5pt
    }
]
coordinates {(Single-o,54.3)(Two-o,44.3)(Three-o,38.0)(Four-o, 28.9)};
\addlegendentry{Hunk-Level}
\addplot[
    color=red!70,
    line width=1pt,
    mark=*,
    mark size=1pt,
    nodes near coords,
    every node near coord/.append style={
        font=\fontsize{6.5}{7.5}\selectfont,
        yshift=-20pt
    }
] 
coordinates {(Single-o,45.0)(Two-o,32.3)(Three-o,30.6)(Four-o,24.1)};
\addlegendentry{w/o FL}
\end{axis}
\end{tikzpicture}
\end{minipage}

\vspace{10pt}

\centering
\vspace{-6mm}
\caption{Repair rate under different bug complexity categories.}
\vspace{-2mm}
\label{fig:complexity_repair_rate}
\end{figure*}
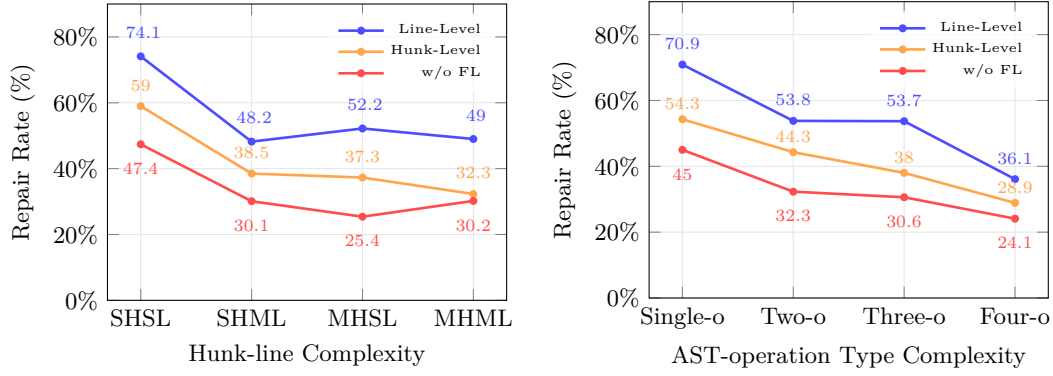
\begin{table*}[t]
\centering
\caption{Logistic regression results of FL effects on repair success under different complexity categories.}
\label{tab:fl_logistic_results_all}

\begin{subtable}{0.46\textwidth}
\centering
\caption{Hunk-line complexity categories}

\scriptsize
\setlength{\tabcolsep}{5pt}
\renewcommand{\arraystretch}{1.02}

\begin{tabular}{lccccc}
\toprule
\multirow{2}{*}{Scope} 
& \multicolumn{2}{c}{Hunk-level FL} 
& \multicolumn{2}{c}{No FL} \\
\cmidrule(lr){2-3}\cmidrule(lr){4-5}
& OR & p 
& OR & p \\
\midrule
Overall  & 0.582 & $<.001$ & 0.399 & $<.001$ \\
SHSL     & 0.502 & $<.001$ & 0.315 & $<.001$ \\
SHML     & 0.672 & 0.037    & 0.462 & $<.001$ \\
MHSL     & 0.544 & 0.084    & 0.311 & 0.002 \\
MHML     & 0.497 & 0.019    & 0.451 & 0.008 \\
\bottomrule
\end{tabular}

\end{subtable}
\hspace{0.01\textwidth}
\begin{subtable}{0.46\textwidth}
\centering
\caption{AST-operation complexity categories}

\scriptsize
\setlength{\tabcolsep}{5pt}
\renewcommand{\arraystretch}{1.02}

\begin{tabular}{lccccc}
\toprule
\multirow{2}{*}{Scope} 
& \multicolumn{2}{c}{Hunk-level FL} 
& \multicolumn{2}{c}{No FL} \\
\cmidrule(lr){2-3}\cmidrule(lr){4-5}
& OR & p 
& OR & p \\
\midrule
Overall    & 0.582 & $<.001$ & 0.399 & $<.001$ \\
single-op  & 0.489 & $<.001$ & 0.336 & $<.001$ \\
two-op     & 0.683 & 0.092    & 0.409 & $<.001$ \\
three-op   & 0.528 & 0.015    & 0.379 & $<.001$ \\
four-op    & 0.719 & 0.321    & 0.561 & 0.092 \\
\bottomrule
\end{tabular}

\end{subtable}
\end{table*}

We use logistic regression~\cite{menard2001applied} to examine the extent to which reduced FL precision affects the likelihood of successful repair. We set line-level FL as the reference category and analyze each complexity level independently. The results are reported using odds ratios (ORs) and corresponding $p$-values. Smaller OR values reflect a greater reduction in the odds of successful repair.

\autoref{fig:complexity_repair_rate} shows that repair rates consistently decrease across all complexity categories as FL becomes less precise (blue vs. orange vs. red). According to \autoref{tab:fl_logistic_results_all}, both hunk-level FL and no-FL settings yield overall $p$-values smaller than 0.001, indicating that FL precision has a statistically significant effect on overall repair effectiveness. 

We further observe that the simplest bugs (SHSL and single-operation bugs) appear sensitive to reductions in FL precision, as both hunk-level FL and no-FL settings produce very low odds ratios, also with $p < 0.001$. In contrast, the highest complexity bugs (four-operation bugs) exhibit the least noticeable repair degradation under hunk-level FL and no-FL settings, as the odds ratios remain relatively higher and the FL degradation effects are not statistically significant ($p = 0.092 > 0.05$).

\textbf{\textit{Finding 5:}} 
\textit{Repair rates consistently decrease as FL becomes less precise across all bug complexity categories, indicating that FL precision remains important regardless of bug complexity. However, as bug complexity increases, repair effectiveness becomes less dependent on precise FL, suggesting it is more likely reliant on structural understanding and reasoning of program logic.}

\begin{tcolorbox}[
    colback=gray!15,
    colframe=gray!15,
    boxrule=0pt,
    arc=2mm,
    left=2mm,
    right=2mm,
    top=1mm,
    bottom=1mm
]
\textbf{RQ2 Summary.}
FL still plays a critical role in the bug repair process. More precise FL not only improves overall repair rates but also helps APR techniques generate more consistent patches. Furthermore, APR techniques with similar repair performance under the same FL strategy can exhibit substantially different effectiveness as FL becomes increasingly imprecise, resulting in progressively larger performance gaps among APR approaches. These results highlight the importance of the stability and generalizability of APR techniques for handling realistic bug-repair scenarios. Moreover, as bug structures become increasingly complex, the relative importance of FL precision decreases, indicating the growing significance of reasoning and program understanding.\end{tcolorbox}

\subsection{APR With Advanced LLMs}
\label{aprwithadvancedllms}

ChatRepair demonstrates overall better performance than CodeCorrector in sections \ref{aprperformancewithdifferentllms} and \ref{buglocalizationeva}, so this study chooses ChatRepair to evaluate performance and cost-efficiency for advanced LLM-based approaches. To reduce API cost, ChatRepair stops when generating the first plausible patch. Subsection \ref{subsec:experimental_design} describes the experimental environment, and \autoref{tab:advancedllmresult} presents the results of ChatRepair combined with four LLMs and two reasoning settings.

Cost analysis shows that inference costs increase substantially when more advanced LLMs are adopted. DeepSeek-V3.2-chat supports approximately 1,565.56 repair attempts per dollar in non-reasoning mode and 119.10 in reasoning mode, whereas DeepSeek-V4-pro decreases to 68.29 in non-reasoning mode, and 9.05 in reasoning mode. GPT-5 supports only 11.19 attempts per dollar in non-reasoning mode and 5.24 attempts per dollar in reasoning mode. These results show that the cost increases considerably as larger and reasoning-oriented LLMs are used. Compared with the advanced LLMs, DeepSeek-V3.2 is much cheaper than GPT-5 and DeepSeek-V4-pro. Therefore, both the choice and reasoning settings of LLMs heavily impact the cost-efficiency.

The performance of different LLMs varies substantially under non-reasoning and reasoning settings. \autoref{fig:advanced_llm_medium_complexity} (a) presents the number of additionally repaired bugs achieved by each LLM when reasoning mode is enabled.

GPT-5 repairs 175 bugs under non-reasoning mode, achieving the highest number of repaired bugs among all evaluated models, while DeepSeek-V4-pro and DeepSeek-V3.2 repair 90 and 75 bugs, respectively. However, when the reasoning effort is increased to high, GPT-5 repairs only 3 additional bugs. In contrast, enabling high-reasoning-mode effort allows DeepSeek-V3.2 and DeepSeek-V4-pro to repair 64 and 81 additional bugs, respectively.

\textbf{\textit{Finding 6:}} \textit{Both LLM selection and reasoning settings have significant impacts on repair effectiveness and cost, while the effectiveness of reasoning strategies differs across LLM families. Enabling reasoning mode provides only limited repair improvement for GPT-5, whereas DeepSeek-V3.2 and DeepSeek-V4-pro benefit considerably more from reasoning-based settings, repairing many bugs that cannot be fixed under non-reasoning configurations.}

\begin{table*}[t]
\caption{The Advanced LLMs Performance}
\label{tab:advancedllmresult}
\centering
\footnotesize
\setlength{\tabcolsep}{3pt}
\renewcommand{\arraystretch}{1.2}

\begin{tabularx}{\textwidth}{l *{5}{>{\centering\arraybackslash}X}} 
\toprule
\multirow{2}{*}{} & Repaired Bugs & Attempts & \mbox{Repaired Bugs / \$}
  & Attempts / \$ \\ 
 \midrule
GPT-5  & 175/211 & 666  & 2.94 & 11.19  \\
GPT-5-reasoning  & 3/36 & 106 & 0.15 & 5.24 \\ \bottomrule
DeepSeek-V3.2-chat  & 75/211 & 1550 & 75.75 & 1565.56 \\
DeepSeek-V3.2-reasoner  & 64/136 & 318 & 23.97 & 119.10 \\ \bottomrule
DeepSeek-V4-pro & 90/211 & 1475 & 4.17 & 68.29 \\ 
DeepSeek-V4-pro-reasoner  & 81/121 & 232  & 3.16  & 9.05  \\ \bottomrule
Llama-3.1-405B  & 49/211 & 1801 & - & - \\ \bottomrule
\end{tabularx}
\end{table*}
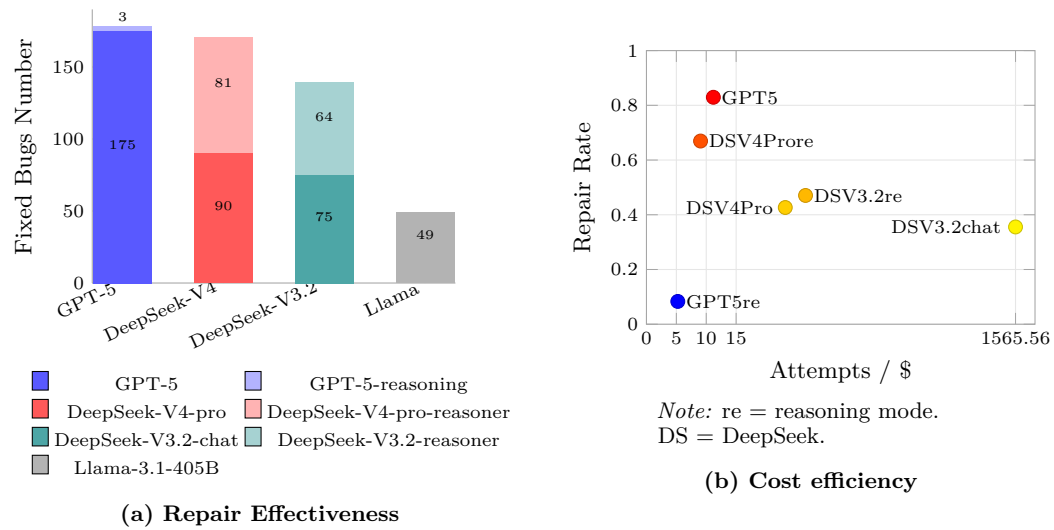
\begin{figure*}[t]
\centering
\footnotesize

\begin{minipage}{0.48\textwidth}
\centering
\begin{tikzpicture}
\begin{axis}[
    ybar stacked,
    width=0.95\columnwidth,
    height=5.2cm,
    ymin=0,
    ymax=190,
    ylabel={Fixed Bugs Number},
    symbolic x coords={
        GPT,
        DeepSeekV4,
        DeepSeekV32,
        Llama
    },
    xtick=data,
    xticklabels={
        GPT-5,
        DeepSeek-V4,
        DeepSeek-V3.2,
        Llama
    },
    xticklabel style={
        rotate=25,
        anchor=east,
        font=\scriptsize
    },
    yticklabel style={font=\scriptsize},
    ylabel style={font=\small},
    bar width=22pt,
    nodes near coords,
    nodes near coords align={vertical},
    every node near coord/.append style={
        font=\tiny,
        black
    },
    legend style={
        at={(0.5,-0.30)},
        anchor=north,
        legend columns=2,
        font=\scriptsize,
        draw=none
    },
    axis x line*=bottom,
    axis y line*=left,
    axis line style={gray!70},
    tick style={gray!70},
    ymajorgrids=false,
    xmajorgrids=false,
]

\addplot[fill=blue!65, draw=none]
coordinates {(GPT,175) (DeepSeekV4,0) (DeepSeekV32,0) (Llama,0)};

\addplot[fill=blue!30, draw=none]
coordinates {(GPT,3) (DeepSeekV4,0) (DeepSeekV32,0) (Llama,0)};

\addplot[fill=red!65, draw=none]
coordinates {(GPT,0) (DeepSeekV4,90) (DeepSeekV32,0) (Llama,0)};

\addplot[fill=red!30, draw=none]
coordinates {(GPT,0) (DeepSeekV4,81) (DeepSeekV32,0) (Llama,0)};

\addplot[fill=teal!70, draw=none]
coordinates {(GPT,0) (DeepSeekV4,0) (DeepSeekV32,75) (Llama,0)};

\addplot[fill=teal!35, draw=none]
coordinates {(GPT,0) (DeepSeekV4,0) (DeepSeekV32,64) (Llama,0)};

\addplot[fill=gray!60, draw=none]
coordinates {(GPT,0) (DeepSeekV4,0) (DeepSeekV32,0) (Llama,49)};

\legend{
GPT-5,
GPT-5-reasoning,
DeepSeek-V4-pro,
DeepSeek-V4-pro-reasoner,
DeepSeek-V3.2-chat,
DeepSeek-V3.2-reasoner,
Llama-3.1-405B
}

\end{axis}
\end{tikzpicture}

\vspace{1mm}
\textbf{(a) Repair Effectiveness}
\end{minipage}
\hfill
\begin{minipage}{0.48\textwidth}
\centering
\begin{tikzpicture}



\begin{axis}[
    scatter,
    only marks,
    width=\textwidth,
    height=5.2cm,
    xmin=0,
    xmax=65,
    ymin=0,
    ymax=1,
    xlabel={Attempts / \$},
    ylabel={Repair Rate},
    xlabel style={font=\small},
    ylabel style={font=\small},
    xtick={
        0,5,10,15,
        61.74
    },
    xticklabels={
        0,5,10,15,
        1565.56
    },
    xticklabel style={font=\scriptsize},
    yticklabel style={font=\scriptsize},
    grid=both,
    major grid style={gray!20},
    axis line style={gray!70},
    tick style={gray!70},
    mark size=2.6pt,
    clip=false,
]

\addplot coordinates {
    (11.19,{175/211})
    (5.24,{3/36})
    (61.74,{75/211})
    (26.61,{64/136})
    (23.22,{90/211})
    (9.05,{81/121})
};

\node[font=\scriptsize, anchor=west]
at (axis cs:11.19,{175/211}) {GPT5};

\node[font=\scriptsize, anchor=west]
at (axis cs:5.24,{3/36}) {GPT5re};

\node[font=\scriptsize, anchor=east]
at (axis cs:61.74-1,{75/211}) {DSV3.2chat};

\node[font=\scriptsize, anchor=west]
at (axis cs:26.61,{64/136}) {DSV3.2re};

\node[font=\scriptsize, anchor=west]
at (axis cs:23.22-18,{90/211}) {DSV4Pro};

\node[font=\scriptsize, anchor=west]
at (axis cs:9.05,{81/121}) {DSV4Prore};

\end{axis}
\end{tikzpicture}

\vspace{1mm}
\footnotesize
\hspace{1.2cm}
\parbox{0.8\linewidth}{
\textit{Note:} re = reasoning mode.\\
DS = DeepSeek.
}
\vspace{1mm}

\textbf{(b) Cost efficiency}
\end{minipage}

\vspace{-2mm}
\caption{Repair effectiveness and cost efficiency of advanced LLMs.}
\vspace{-2mm}
\label{fig:advanced_llm_medium_complexity}
\end{figure*}

\autoref{fig:advanced_llm_medium_complexity} (b) presents the repair rates of different LLMs and reasoning settings under the cost metric of attempts per dollar (Attempts/\$). Combined with \autoref{fig:advanced_llm_medium_complexity} (a), GPT-5 repairs 175 bugs under the non-reasoning setting, which is higher than all other evaluated LLMs under both reasoning and non-reasoning settings (171 for DeepSeek-V4-pro and 139 for DeepSeek-V3.2). Moreover, GPT-5 under the non-reasoning setting is even cheaper than DeepSeek-V4-pro under the reasoning setting in terms of attempts per dollar (11.19 vs. 9.05). Although GPT-5 under the non-reasoning setting is more expensive than DeepSeek-V3.2 under the reasoning setting, the repair rate is markedly higher.

\autoref{fig:advanced_llm_medium_complexity} (b) shows that under the non-reasoning setting, DeepSeek-V3.2 achieves competitive repair performance compared with DeepSeek-V4-pro, while maintaining a substantially lower cost. Therefore, using DeepSeek-V3.2 in non-reasoning mode for overall bug repair, and subsequently applying GPT-5 in non-reasoning mode to repair the remaining unresolved bugs, provides a more effective trade-off between repair performance and cost.

\textbf{\textit{Finding 7:}} \textit{GPT-5 achieves the strongest overall repair effectiveness, whereas DeepSeek-V3.2-chat demonstrates significantly better cost-efficiency. Combining cost-efficient LLMs with more advanced LLMs for unresolved bugs can therefore achieve a better balance between repair effectiveness and cost-efficiency in LLM-based APR techniques.}

\begin{table*}[t]
\caption{Repair performance (Top-$k$) across APR configurations.
Values are reported as incremental fixes and cumulative total in parentheses.}
\label{tab:pass_at_k_performance}
\centering
\scriptsize
\setlength{\tabcolsep}{5pt}
\renewcommand{\arraystretch}{1.10}

\begin{tabular}{lllcccccc}
\toprule
\textbf{APR} & \textbf{LLMs} & \textbf{FL} &
\textbf{Top-1} &
\textbf{Top-2} &
\textbf{Top-3} &
\textbf{Top-4} &
\textbf{Top-5} &
\textbf{Top-10} \\
\midrule

\multirow{5}{*}{CC}
& L3.1-8B                &ll          & 8   & 2 (10)   & 1 (11)   & 0 (11)   & 0 (11)   & 0 (11)   \\
& 4o-mini                 &ll          & 80  & 23 (103) & 8 (111)  & 6 (117)  & 4 (121)  & 5 (126)  \\
& DS-V3.2               &ll          & 191 & 49 (240) & 18 (258) & 14 (272) & 6 (278)  & 17 (295) \\
& DS-V3.2               &hl          & 106 & 32 (138) & 12 (150) & 12 (162) & 2 (164)  & 15 (179) \\
& DS-V3.2               &w/o FL              & 80  & 15 (95)  & 6 (101)  & 1 (102)  & 2 (104)  & 8 (112)  \\
\midrule

\multirow{5}{*}{CR}
& L3.1-8B                &ll          & 15  & 5 (20)   & 1 (21)   & 9 (30)   & 6 (36)   & 14 (50)  \\
& 4o-mini                 &ll          & 67  & 16 (83)  & 9 (92)   & 14 (106) & 9 (115)  & 25 (140) \\
& DS-V3.2               &ll          & 194 & 30 (224) & 8 (232)  & 19 (251) & 8 (259)  & 33 (292) \\
& DS-V3.2               &hl          & 93  & 38 (131) & 22 (153) & 26 (179) & 16 (195) & 42 (237) \\
& DS-V3.2               &w/o FL              & 128 & 30 (158) & 14 (172) & 11 (183) & 7 (190)  & 15 (205) \\
\midrule

\multirow{7}{*}{CR adv.LLM}
& L3.1-405B               &ll       & 15  & 12 (27)  & 0 (27)   & 5 (32)   & 3 (35)   & 14 (49)  \\
& GPT-5                        &ll       & 103 & 43 (146) & 16 (162) & 4 (166)  & 5 (171)  & 4 (175) \\
& DS-V3.2                &ll       & 45  & 8 (53)   & 4 (57)   & 3 (60)   & 3 (63)   & 12 (75)  \\
& DS-V4-pro                  &ll       & 31  & 26 (57)  & 5 (62)   & 5 (67)   & 7 (74)   & 16 (90)  \\
& GPT-5(re)            &ll       & 1   & 0 (1)    & 2 (3)    & --       & --       & --       \\
& DS-V3.2(re)    &ll       & 33  & 19 (52)  & 12 (64)  & --       & --       & --       \\
& DS-V4-pro(re)     &ll       & 58  & 15 (73)  & 8 (81)   & --       & --       & --       \\
\bottomrule

\end{tabular}

\vspace{1mm}

\parbox{\textwidth}{

\raggedright

\scriptsize

\textit{Note:} CC=CodeCorrector, CR=ChatRepair, L3.1 = Llama-3.1, 4o-mini = GPT-4o-mini, DS = DeepSeek, re = reasoning, w/o = without, ll = line-level FL, hl = hunk-level FL.}

\vspace{-4mm}

\end{table*}

\autoref{tab:pass_at_k_performance} presents the $\textit{Top}-k$ metric for both APR techniques across three LLM families and three different FL strategies. The first five repair rounds usually resolve the largest proportion of bugs across evaluated APR techniques, LLMs, and FL settings. However, the marginal repair gain decreases considerably as additional repair rounds are introduced, with later rounds contributing only a limited number of newly repaired bugs.

\textbf{\textit{Finding 8:}} \textit{The majority of bugs are repaired within the first five repair rounds, whereas the marginal repair effectiveness decreases significantly in later rounds.}

\begin{tcolorbox}[
    colback=gray!15,
    colframe=gray!15,
    boxrule=0pt,
    arc=2mm,
    left=2mm,
    right=2mm,
    top=1mm,
    bottom=1mm
]

\textbf{RQ3 Summary.}
Using advanced LLMs and reasoning modes improves repair rates and enables APR techniques to fix more structurally complex bugs. However, these improvements come with substantially higher costs, making the trade-off between repair performance and cost increasingly important. Moreover, the benefits of reasoning modes vary across different LLM families. In our experiments, GPT-5 shows the smallest gain from reasoning, whereas the DeepSeek series repairs many additional bugs that cannot be fixed in non-reasoning mode. We further find that applying cost-efficient LLMs first, and then using advanced LLMs to repair the remaining unresolved bugs, can achieve a better balance between repair effectiveness and cost.
\end{tcolorbox}

\section{Discussion}
We investigated the effectiveness of LLM-based APR under different bug complexity categories and fault localization strategies. The results reveal several insights regarding the repair effectiveness and cost-efficiency of the evaluated LLM-based APR techniques, providing useful guidance for future APR research and practical deployment.

\textbf{Impacts of bug complexity}: Our results show that bug complexity substantially affects the effectiveness of LLM-based APR techniques. The simplest bugs achieve much higher repair rates, while structurally complex bugs containing multiple buggy lines or diverse AST edit operations remain more challenging.

The repair performance across different bug complexity levels suggests that future APR evaluations could pay more attention to complex repair scenarios rather than to single-line or single-operation bugs. Structurally complex bugs may better reflect some of the challenges encountered in practical APR tasks. New benchmarks and standard performance metrics could consider evaluations across different bug categories, domains, and complexity levels to better characterize tool performance~\cite{dikici2025advancements, puvvadi2025coding, renzullo2025automated}. 

\textbf{Impacts of fault localization}: Precise FL not only improves repair effectiveness, but also helps APR techniques generate patches that are more consistent with the original program logic. Such consistency may facilitate the understanding, validation, and maintenance of repaired programs.

Additionally, existing LLM-based APR methods are affected differently by less precise FL information. This difference highlights robustness to imprecise FL as an important requirement for practical APR techniques~\cite{dikici2025advancements}. Since real-world FL is often noisy and incomplete, methods that maintain stable repair performance when the precision of FL decreases are more likely to be effective in practice. For complex bugs, the relatively smaller effect of FL precision indicates that FL may not be the only bottleneck. Repairing these bugs may also depend on reasoning ability and deeper program-understanding capabilities.

\textbf{The Trade-off Between Repair Effectiveness and Cost-efficiency}: Using more advanced LLMs and reasoning modes substantially increases computational cost, while often improving repair effectiveness. Based on the results presented in section \ref{aprwithadvancedllms}, GPT-5 achieves the highest repair effectiveness, while DeepSeek-V3.2-chat provides the most cost-efficient repair performance. Our results suggest that such a two-stage strategy could be a promising direction for future investigation.

The marginal repair efficiency drops as the number of repair rounds increases~\cite{palavalli2024using, zhao2024repair}, which further impacts overall cost, especially when using advanced LLMs such as GPT-5 and DeepSeek-V4-pro in reasoning mode. Our results also suggest that selecting an appropriate maximum number of repair rounds is important for cost-efficiency.

\section{Threats to Validity}
\textbf{External Validity}: Our dataset is collected solely from the competitive programming platform AtCoder and focuses only on C++ programs, which are typically shorter but more algorithm-oriented~\cite{lin2017quixbugs, le2015manybugs} than industrial software systems. Thus, the findings of our study may not generalize fully to real-world software projects with larger codebases, complex dependencies, and practical software engineering constraints. Furthermore, we evaluate only two APR techniques and three LLM series, and thus, the best-performing approaches identified in this study should not be regarded as definitive state-of-the-art APR solutions.

The evaluated LLMs have different knowledge cutoff dates, and we collect benchmark data after 2025-10-01, which is after the knowledge cutoff date of the GPT and Llama models we used. DeepSeek did not publish its knowledge cutoff date, so it remains difficult to completely rule out the possibility that some competitive programming solutions or related code patterns were exposed during model pretraining~\cite{ni2025training, ramos2025large}.

\textbf{Internal Validity}: We set the temperature to 1.0 for all evaluated LLMs and used high reasoning-effort settings instead of the maximum settings for GPT-5, DeepSeek-V3.2-reasoner, and DeepSeek-V4-pro due to their substantial time costs. Different parameter configurations and reasoning settings may lead to variations in repair performance. In addition, API-based LLMs may produce nondeterministic outputs across different runs, which could also affect repair results. Furthermore, different prompting strategies, repair rounds, and stopping criteria may influence the effectiveness and cost-efficiency of APR.

\textbf{Construct Validity}: We determine repaired bugs mainly based on whether the generated patches can pass all collected test cases. However, the test cases may not be sufficiently comprehensive to fully validate the semantic correctness of repaired programs. Therefore, some plausible patches may still contain hidden defects or overfitting issues.

\section{Conclusion and Future Work}
This work conducts an empirical study on how bug complexity and FL affect LLM-based APR techniques. We utilize Git, srcML, and GumTreeDiff to categorize bug complexity from both hunk-line-level and AST-operation-level perspectives, and to evaluate APR effectiveness under accurate, vague, and no fault-localization settings. 

The experimental results show that structurally complex bugs and imprecise FL often reduce repair effectiveness. Nevertheless, well-designed LLM-based APR techniques still maintain considerable repair rates under these challenging conditions. Furthermore, different LLMs exhibit clear trade-offs between repair effectiveness and cost-efficiency, highlighting the importance of balancing repair effectiveness and cost for practical APR tools.

Future work may develop a benchmark that can automatically collect, filter, and categorize bug complexity. Also, we will consider practical factors such as API response latency~\cite{park2026minimizing, liu2026bag} and overall inference time when evaluating the efficiency of LLM-based APR systems. We plan to explore additional LLM families, such as Gemini and Claude, to further evaluate the effectiveness of LLM-based APR. Finally, the agent-based APR frameworks should be further investigated, as agent-based techniques have become increasingly popular across various software engineering and reasoning tasks in recent years.

\section{Data Availability}
The dataset and source code used in this study are publicly available at the following links:

Dataset: https://doi.org/10.5281/zenodo.20257267

Code: https://github.com/Leo6-sys/Rethink-APR-Empirical-Study-2026

\bibliography{References}

\end{document}